\documentstyle[epsf]{mn}

\title{Periodic Microvariation of B416, a New LBV in M33}
\author[O. Shemmer, E. M. Leibowitz and P. Szkody]
       {Ohad Shemmer,{\Huge$^{1}$}\thanks{ohad@wise.tau.ac.il;
        elia@wise.tau.ac.il;szkody@alicar.astro.washington.edu}
       Elia M. Leibowitz{\Huge$^{1}$}{\Huge$^\star$} and Paula Szkody
       {\Huge$^{2}$}{\Huge$^\star$} \\
       {$^{1}$}School of Physics and Astronomy and the Wise Observatory, The
Raymond and Beverly Sackler Faculty of Exact Sciences, \\ Tel-Aviv
University, Tel-Aviv 69978, Israel \\ {$^{2}$}Department of Astronomy,
University of Washington, Seatle, WA 98195, \\ Based on observations with the
Apache Point Observatory (APO) 3.5m telescope, \\ which is owned and operated by
the Astrophysical Research Consortium (ARC)}

\date{Accepted 1999 September 2.
      Received 1999 August 15;
      in original form \today}

\pagerange{\pageref{firstpage}--\pageref{lastpage}}
\pubyear{1999}

\begin{document}

\maketitle

\label{firstpage}

\begin{abstract}

We report optical photometric and spectroscopic observations of the M33
star B416, establishing it as the fifth identified LBV in that galaxy.
The B magnitude of the star varies with a period of 8\fd26 and an
amplitude of 0.03 magnitude. The red end of the optical spectrum does
not seem to partake in these oscillations, which we identify as the
microvariations, known to persist in LBV stars. B416 is the first LBV
with such oscillations that are strictly periodic and coherent over at
least 10 years. The brightness of the star is also varying on a longer
(years) time scale, again mainly in its blue light. The spectrum has
most of the characteristics of an LBV in quiescence, showing Balmer, He
{\sc{i}}, Fe {\sc{ii}} and [Fe {\sc{ii}}] emission lines. The flux and
profile of some of these lines are varying but the scarcity of the
observations prevents us from establishing their characteristic
time scale. B416 is at the centre of an H {\sc{ii}} region, another
indication of the relatively young age of this stellar system.

\end{abstract}


\section{Introduction \label{1}}
\subsection{The M33 survey \label{1.1}}

During the years 1986 and 1987 the local group galaxy M33 was monitored
nightly at the Tel-Aviv University Wise Observatory (WO) for two months
in each year. This program was intended to search for potential SS433
candidates in M33, since this almost face on nearby galaxy is a most
promising case for locating such a unique system outside the Milky
Way.  The survey comprised 15 star fields in M33, with each one
containing one or more extremely bright blue and red giants selected
from the luminous star survey performed earlier by Humphreys \& Sandage
(1980).  Our primary goal was to find luminous stars in M33 that show
large variability in their light curves.  We chose to check for
periodic variations all stars that had a standard deviation larger than
0.03 mag in their measured magnitudes.  Periodic oscillations might
hint at binarity and would make such a luminous star an SS433 candidate
which could be confirmed by a search for "moving lines" in its spectrum
\cite{mar79}.

The reduction procedure of the survey images (described briefly in
\S~\ref{2.1}) yielded 12 objects with measurable light variations. For
one of these objects a period of $\sim$8 days was found \cite{she98}.
This object was identified as B416, a blue giant in the M33 survey of
bright blue and red stars \cite{hum80}, as an H$\alpha$ emitter in M33,
no.  24, 130e of field l in Calzetti et al. (1995), who also give the
1950.0 coordinates of the star as RA: 01:31:17.370 DEC: +30:26:26.26.
We reobserved the star in 1997/98, photometrically and spectroscopically,
in order to reconfirm its periodic variability and to check the possibility
regarding it as an SS433 candidate. The collection of photometric and
spectroscopic data does not support the SS433 hypothesis. It led us instead,
to accept the identification of the star as an LBV, and to interpret all our
results in the light of this identification.

\subsection{B416 \label{1.2}}

B416 has been recognized as a bright blue star in several surveys in the past,
but no variability of it has been so far reported.  Humphreys \&
Sandage (1980) list it as a bright blue star with m$_V$=16.3 and
B$-$V=0.29 while according to Calzetti et al. (1995) the star's
magnitude is m$_V$=16.76 and it is among the stronger H$\alpha$
emitters in M33 with an EW(H$\alpha$) of 109.1\AA. According to Massey
et al. (1996), B416 is also one of the brightest UV sources in M33,
listed there as UIT301 and is classified as one of five new LBV
candidates in that galaxy. The following photometric properties of B416
are also given there: m$_V$=16.29, B$-$V=0.11, M$_V=-$8.8 and
U$-$B=$-$0.82, indicating again a very luminous object. The
different magnitude estimates given to B416 by different authors might
hint on its variability. The discrepancy between the B$-$V colours
may be related to genuine colour variations (see \S~\ref{3}). It may also be due
to differences in the assumed colour correction that should be applied to this
star, since the reddening across M33 is not uniform \cite{mas95}.

\begin{figure}
\centerline{\epsfxsize=3.3in\epsfbox{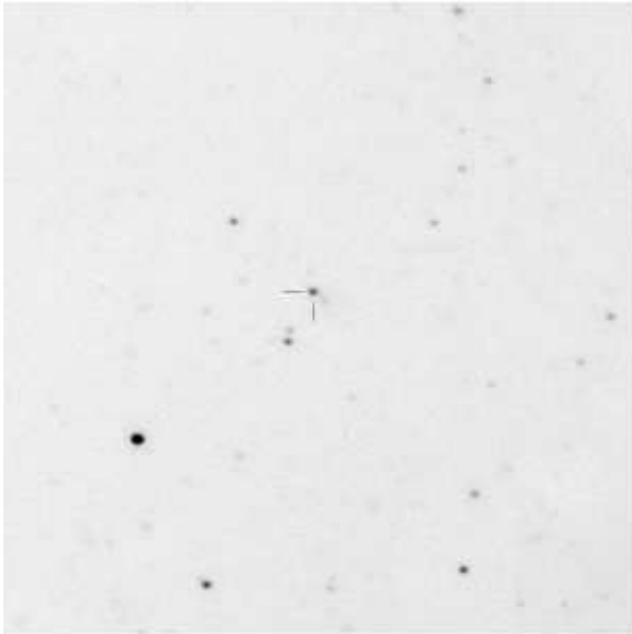}}
\caption{Finding chart of B416 in V. North is up and East is left in this
3.5$\arcmin$ X 3.5 $\arcmin$ field.}
\label{B416find}
\end{figure}

\begin{figure}
\centerline{\epsfxsize=3.3in\epsfbox{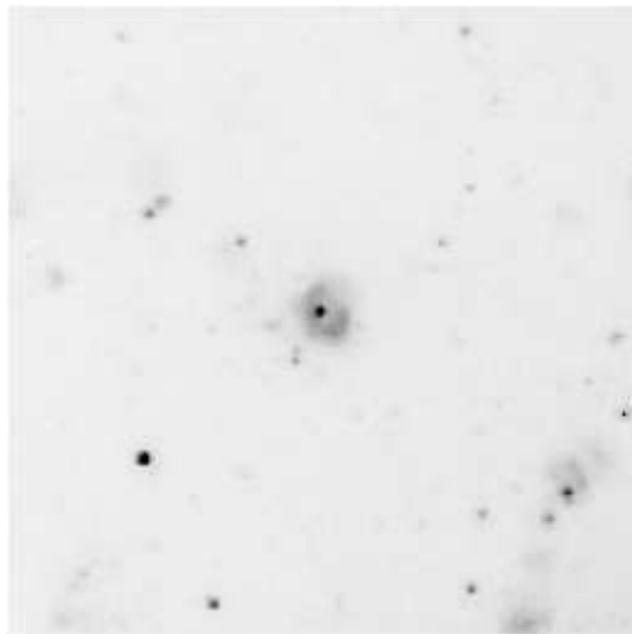}}
\caption{B416 inside a ring-shaped H {\sc{ii}} region in this H$\alpha$ image
of the same field seen in Figure \protect{\ref{B416find}}.}
\label{nebula}
\end{figure}

\section{Observations and Reductions \label{2}}
\subsection{Photometry \label{2.1}}

Photometry of B416 was performed with the 1m telescope of the WO. The
survey observations in 1986 \& 1987 used the RCA $320\times 520$ pixel
thinned CCD camera and were performed with no filter (Clear). Following
the discovery of periodicity in the light curve of B416, we renewed in
1997 the nightly monitoring program of its field, this time using the
Tektronix $1024\times 1024$ pixel back-illuminated CCD camera,
described in Kaspi et al. (1995). Observations in the 1997-98 season
were carried out in B \& V filters as well as in Clear and the latest
observational season in 1998-99 used mainly U,B,V,R,I filters and few
Clear observations. The I filter used at the WO is somewhat redder than
the standard Johnson bandpass, because its red end is determined by the
CCD camera cut-off, which has a longer tail toward  lower frequencies.
In order to look for shorter time scale variations (i.e.  minutes or
hours) we devoted three nights to conducting continuous time series
measurements of B416.  On the first night, 1997 November 26, we took 46
Clear frames during 3 hours and on the next two nights, 1998 December 6
and 7 we took 63 images spanning 8 hours and 48 images spanning 6 hours
respectively, using filters B,R \& I sequentially for both nights. The
total number of data points taken in each filter every season appears
in Table~\ref{tbl-1}. Several attempts were made to calibrate the
magnitude of B416. They all failed due to unfortunate non photometric
sky conditions.  Throughout this work we therefore use instrumental,
relative magnitudes, which are far less sensitive to these conditions.

\begin{table*}
\centering
\begin{minipage}{140mm}
\caption{Number of photometric data points. \label{tbl-1}}
\begin{tabular}{@{}lcccccc@{}}
Season/Filter & Clear & U & B & V & R & I \\
1986 & 25 & -- & -- & -- & -- & -- \\
1987 & 43 & -- & -- & -- & -- & -- \\
1997-98 & 96 & -- & 29 & 30 & -- & -- \\
1998-99 & 8 & 30 & 76 & 40 & 77 & 72
\end{tabular}
\end{minipage}
\end{table*}

Aperture photometry measurements on the biased and flat field corrected
CCD frames were carried out using the NOAO IRAF\footnote{{IRAF (Image
Reduction and Analysis Facility) is distributed by the National Optical
Astronomy Observatories, which are operated by AURA, Inc., under
cooperative agreement with the National Science Foundation.}}
{\sc{daophot}} package \cite{ste87}. A final set of data files was
later obtained using the WO {\sc{daostat}} reduction program, which
determined the object star magnitude, uncertain to within $\pm$15
mmag,  relative to a set of standard stars in its field (see Netzer et
al. (1996) for a full description of the reduction process). Such a set
was created for each filter. The 1986 data, that were acquired during
the initializing run of the first CCD camera of the WO, turned out to
be too noisy for our time series analysis (uncertainty in magnitude of
over 30 mmag) and were left out of our further analysis.

\subsection{Spectroscopy \label{2.2}}

Spectroscopy of B416 was carried out in the 1998-99 season in order to
look for SS433-like features in the spectrum of the star. The
observations were done at the WO and at the Apache Point Observatory
(APO) in New Mexico using the 3.5m telescope. Spectra were taken on
seven nights at the WO and on three nights at the APO. The WO images
were taken with the Faint Object Spectroscopic Camera (FOSC) attached
to the Tektronix CCD described above using a 10\arcsec \ wide blue
sensitive long-slit and a 600 lines/mm grating ~\cite{kas95}. Each
night four images of B416 were taken, two at PA=45\degr \ and two at
PA=135\degr \ in order to gather more information on the morphology of
the nebula surrounding the star.  B416 lies at the detection limit of
our spectrograph, which is $\sim$16 mag. We therefore took two images
of the star for each PA, each exposed for 1 hour, thus totaling in 4
hours of observation each night. Since our FOSC imager setup has a
resolution of 2\arcsec \ per pixel, we were not able to resolve B416
from its SE neighbouring star B417 (UIT299 in Massey et al. 1996) and
from the surrounding nebulosity. The APO observations were performed on
1998 August 31, November 24 and December 14 UT using the double imaging
spectrograph (DIS) attached to the 3.5m telescope. Blue and red
sensitive chips were used each night where the blue chip is a 512x512
UV coated Tek CCD and the red is an 800x800 TI CCD. The slit was
1.5\arcsec \ wide oriented in an E$-$W direction and the gratings were
830.8 (1200) lines/mm in blue (red). These observations were made
during times of fairly good seeing (1.5, 1.2 and 1.7 arcsec
chronologically on each of the 3 nights) so that B416 is well resolved
from its neighbour B417. Some blend with nebular emission from
foreground, background and nearby nebulosity is of course unavoidable.

Reduction of all spectra was conducted using the IRAF {\sc{specred}}
package. Since no spectroscopic standards were observed at the WO when
the spectra were taken, the spectrum was flux calibrated using the WO
standard coefficients, which generally do not change significantly from
night to night. The spectra extend from 3600 to 7200\AA \ with a
dispersion of 3.8\AA \ per pixel (about 8\AA \ resolution). The APO
spectra on the other hand were taken along with spectroscopic standards
and were therefore flux calibrated accordingly.  The APO images of the
spectrum came in pairs of red and blue bands, where the blue images
extend from 4200 to 5000\AA \ with a dispersion of 1.6\AA \ per pixel
(about 3\AA \ resolution) and the red images extend from 5800 to
6800\AA \ with a dispersion of 1.3\AA \ per pixel (about 2.5\AA
\ resolution). In both bands we were able to spatially resolve some of
the nebulosity around the star (this was impossible with the images
taken at WO), and were therefore able to extract also the spectrum of
the line-of-sight nebulosity in the following way: the images of B416
were reduced once using a sky background that was subtracted from the
star's aperture and once using the background that was selected as the
mean spatial extension of the nebula seen in the images. We therefore
obtained two types of spectra for each band: one including the light
from the star$+$the line-of-sight nebulosity and another including the
star's light only (assuming a spatially homogeneous nebula around the
star). Subtracting the latter spectra from the first resulted in a
spectrum of the line-of-sight nebula alone. The fact that the nebula
surrounding B416 is in fact not homogeneous but instead has a ring-like
structure can be seen in Fig.~\ref{nebula}. All spectra were cleaned of
cosmic rays and bad pixels.

\section{Data Analysis \label{3}}

\subsection{The LC and periodicity\label{3.1}}

\subsubsection{Periodic Variation\label{3.1.1}}

\begin{figure}
\centerline{\epsfxsize=3.3in\epsfbox{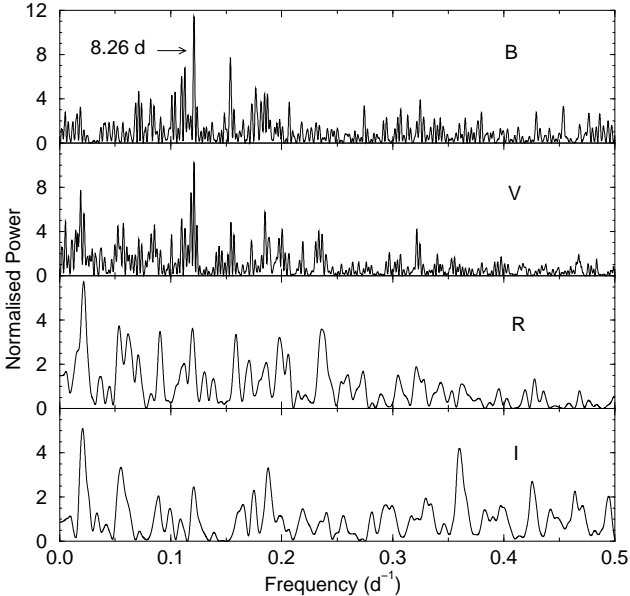}}
\caption{PS of the BVRI data from top to bottom. Note the decreasing power of
the 8\fd26$\pm$0\fd03 period with the increase in wavelength.}
\label{BVRIps}
\end{figure}

The normalised power spectrum (PS) \cite{sca82} of the BVRI data is
plotted in Fig.~\ref{BVRIps}. The B \& V PS imply a period of
8\fd26$\pm$0\fd09, which can also be seen in R \& I, but with a very
low power. This period is also seen in the Clear PS, although due to
some high noise in this data set, a higher peak in
the PS is seen for 8\fd14$\pm$0\fd03 \cite{she98}. This peak is however still
consistent with the above period within the uncertainty limits. The scarcity of
the U data (only 30 data points) made it unusable for period searching. The
8\fd26$\pm$0\fd03 period in the last two seasons is within 2$\sigma$ of
the 8\fd6$\pm$0\fd2 period found for Clear in 1987 \cite{she98} thus
indicating that the period remained stable during the past decade. We
remark that when calculating the PSa, for each of the 3 nights that
were used to look for shorter periodicities, we utilized only one data point
randomly selected, in order to keep even weights in the data.

The reliability of the period in B, V and Clear was checked by
splitting each data set into two seasonal subsets and verifying that
the same period (within the uncertainty limits, described below)
appears in every subset. We checked the significance of the period by a
Bootstrap technique \cite{efr93} on a sample of 1000 artificial LCs
constructed by randomly permutating the magnitudes of each subset. The
reality of the periodicity found in the B, V and Clear filters was
found to be significant at a 99\% confidence level. The uncertainty in
the value of the period was determined again by the Bootstrap
technique, this time by a least square fitting of a sine function to
the data, subtracting it from the LC and then randomly adding the
residuals to the sine data values. Next we found the peak frequency of
1000 artificial LCs so created. In the histogram of the distribution of
the peak frequencies we determined the interval that contains 980
results. This is given above as a 98\% significant interval around the
period value.

In order to check whether B416 varies on a time scale shorter than
days, we analysed the data collected during the three nights of
continuous monitoring.  All three nights did not show any periodicity
nor any significant variability, i.e. above 15 mmag.

\subsubsection{Long Term Variations\label{3.1.2}}

\begin{figure}
\centerline{\epsfxsize=3.3in\epsfbox{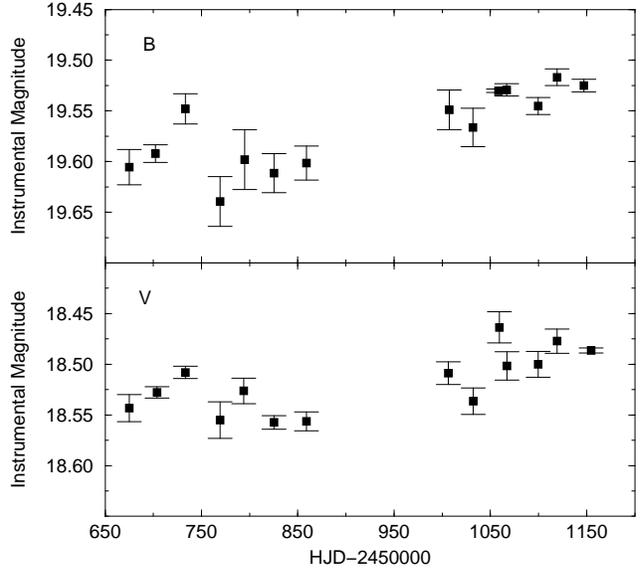}}
\caption{The LCs of B and V binned into roughly monthly bins. The gap separates
between the 1997-98 and 1998-99 seasons. Note the gradual increase in brightness, which is more
prominent in B than in V.}
\label{BVLCbin}
\end{figure}

When we compare the magnitude of B416 in Clear between 1987 and 1997,
we discover that in 1997 the star was fainter by 0.17$\pm$0.03 mag than
in 1987. This comparison was done using three standard stars in the
field of B416, which did not vary by more than 0.02 mag during the past
11 years. We also measured the mean flux centred at the $\lambda$6100
continuum of our APO spectra and used it to calibrate the star's V
magnitude using the value of F$_{\lambda}$ at that wavelength given by
Calzetti et al. (1995); we obtained a V magnitude of 16.7$\pm$0.1 mag
for B416, which is consistent with the value 16.76 (see \S 3.3) they
give based on their observations of M33 during the 1986 \& 1987
seasons. It therefore seems that the red continuum of B416 remained
essentially constant between the epochs of 1987 and 1998. The fading in
Clear is accordingly due to a fading in the blue band of more than 0.17
mag, since Clear covers the whole range of optical light and our CCD is
more sensitive in the red than in the blue band. Another change in the
B and V brightness of the star is apparent between the 1997-98 and
1998-99 seasons. In order to amplify the effect we binned the data into
7 bins in each season (roughly one month per bin) and then calculated
the mean magnitude of each binned season.  Fig.~\ref{BVLCbin} displays
the binned B and V LCs. One can see that in B the star brightened by
0.06 mag and in V it brightened by 0.04 mag between the two seasons.
In the B case this brightening is significant, since the difference in
the mean magnitudes is larger than the combined standard deviation of
each season's mean, while in V, the uncertainty in the magnitude is a
little larger than the difference in mean magnitudes.  This again
manifests the fact that much like the 8\fd26 periodic oscillations, the
longer term variations are also occuring mainly in the blue end of the
spectrum.

\subsection{The shape of the periodic modulation \label{3.2}}

\begin{figure}
\centerline{\epsfxsize=3.3in\epsfbox{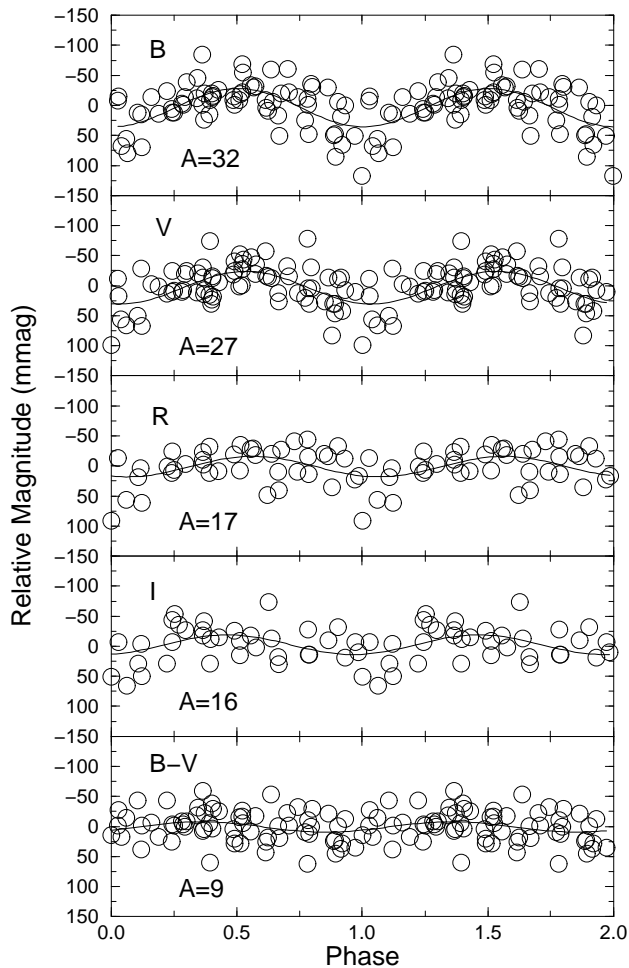}}
\caption{BVRI and B$-$V data points folded onto the 8\fd26 period. The
amplitude (A) of a sine function (solid line) fitted by least squares to each
data set is given in mmag below each set. The symbol's radius is approximately
the error in magnitude.}
\label{LCfold}
\end{figure}

\begin{figure}
\centerline{\epsfxsize=3.3in\epsfbox{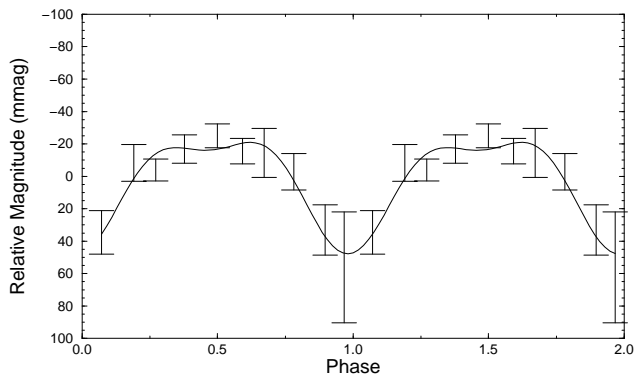}}
\caption{Folded B LC of B416 binned into 0.1 phase bins. A two harmonic fit
with the 8\fd26 period is superposed on the data. The magnitude error bars
indicate the STD of the mean magnitudes in a bin.}
\label{harmony}
\end{figure}

The calculated ephemeris for the B \& V filters is:

T({\it{min}})$=$HJD $2450672.2(\pm1.6)+8.26(\pm0.03)$ E \\
The set of folded LCs along with a fitted sinusoid are plotted in
Fig.~\ref{LCfold} and a plot of a binned B LC along with a two-harmonic
fit is shown in Fig.~\ref{harmony}. It is apparent in Fig.~\ref{LCfold}
that the amplitude of the modulation decreases with
increasing wavelength.  In fact the modulation is clearly
recognizable only in B \& V, while in R \& I the magnitude is practically
constant.  This fact also follows from the PS of each filter's data set
(Fig.~\ref{BVRIps}). Since B \& V are varying in phase, the modulation
of the B$-$V colour has only a small amplitude.

In Figure~\ref{harmony} one can see that the folded B LC,
the filter in which the signal is strongest, is not a perfect
sinusoid. This pattern of a flat-top LC is also apparent in the folded
V and Clear LCs, although not as vivid as in B. In order to verify
the significance of the distorted sinusoid, we calculated the phase of
each filter's second harmonic. They all fell within 0.13 of the cycle
of one another. For random deviation from a sine function, the
probability of obtaining such a crowding of the phase values is 1.7\%.
Thus the non harmonic structure of the periodic LC seems to be
systematic and significant.

\subsection{Spectroscopy \label{3.3}}

\subsubsection{The spectra of B416 \label{3.3.1}}

\begin{figure}
\centerline{\epsfxsize=3.3in\epsfbox{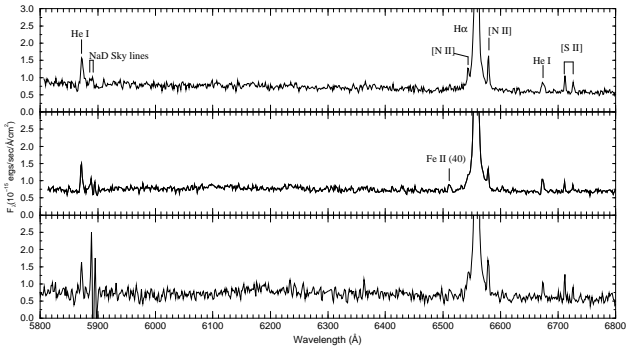}}
\caption{Red band spectrum of B416 and the surrounding nebula.
    {\it{Top to bottom:}} August, November and December 1998. The major stellar
    and nebular lines are identified (the NaD lines are residuals from the
    subtraction of the strong night-sky lines).}
\label{APOred}
\end{figure}

\begin{figure}
\centerline{\epsfxsize=3.3in\epsfbox{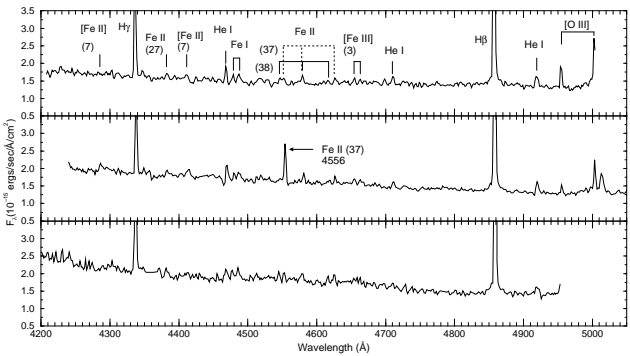}}
\caption{Blue band spectrum of B416 and the surrounding nebula. {\it{Top to bottom:}} August, November and December 1998. The major stellar and nebular lines are identified. Note the dramatic flux increase in the Fe {\sc{ii}} $\lambda$4556 line in November.}
\label{APOblue}
\end{figure}

We did not perform any spectral measurements on the WO spectra and
restricted ourselves only to line identification and general properties
of the spectrum due to the poor resolution, S/N ratio and inability to
resolve B416 from B417 and the nebulosity. The APO spectra on the other
hand were clean of large light contributions from external sources
(except for several faint stars around B416) and we were able to
separate B416 from the surrounding nebula. The three sets of APO
spectra were analysed using the IRAF {\sc {splot}} task to measure line
and continuum properties.  The relatively high resolution of these
spectra enabled us to deblend sets of nearby lines and to detect
variations in line profiles.  Table~\ref{lines} lists the mean measured
properties of some of the major identified lines in the spectra, which
are presented in Figures~\ref{APOred}\& \ref{APOblue}. The line
properties such as flux, EW etc. were averaged among the three spectra
and the errors in the measured properties represent the STD of the
mean. In this sense an error larger than about 10\% (a rough estimation
of the uncertainty in flux calibration) in a measured property in one
of the major strong lines of the star indicates a possible variation
between the three epochs. In the nebular lines the errors are much
larger due to lower fluxes and differences in the subtraction
parameters used to extract the line-of-sight nebula between the three
epochs (see \S~\ref{2.2}).

The spectrum of B416 in general resembles that of a blue hypergiant, or
more specifically an LBV in quiescence, as suggested in Massey et al.
(1996). It shows mainly Balmer, He {\sc {i}}, Fe {\sc {ii}} and [Fe
{\sc {ii}}] emission (B416 is also identified in Corral (1996) as
object S145 where it is classified as a compact H {\sc {ii}} region).
The main spectral feature is the strong H$\alpha$ emission, which can
be evaluated from Table~\ref{lines} as having the luminosity of
$\sim$1700 L$_{\sun}$ (using a distance of 800kpc to M33) and an EW of
-106$\pm$2 \AA , which is consistent with the -109.1 \AA \ value
measured for this star by Calzetti et al. (1995).  The measured
wavelengths of the centroids of the main emission lines do not vary by
more than $\sim$0.5 \AA \ (smaller than our resolution) from night to
night and remain blueshifted by about $\sim$200 km/sec, which is
characteristic of the systemic velocity of M33. The structure of the
Balmer and some of the He {\sc {i}} emission lines has a rather complex
form and might decompose into a narrow component sitting on top of a
broad emission \cite{smi97}. In particular the H$_{\alpha}$ and
H$_{\beta}$ lines have wings with width of the order of 1000 km/s,
whereas the half width at half maximum of these lines is of the order
of 150 km/sec.

In Figure~\ref{APOred} one can see that the red-band continuum remained
practically constant during the three observations, while in
Figure~\ref{APOblue} the blue-band continuum shows a tendency of increasing
flux towards shorter wavelengths, chronologically from August to December 1998.
Also apparent in Figures~\ref{APOred}\& \ref{APOblue} and in Table~\ref{lines}
is the fact that the Balmer emission remains constant to within $\sim$10\%
accuracy in flux and profile, while the He {\sc {i}} emission lines vary in
flux and shape. In the November spectrum we see the development of P Cygni
profiles in the He {\sc {i}} emission lines, which is especially pronounced in
the He {\sc {i}} $\lambda$4471 line.  These P Cygni profiles in the He {\sc
{i}} lines indicate velocities of $\sim$300 km/sec as measured from the peak of
emission to the minimum of the P Cygni trough. Another feature apparent in the
November spectrum is the strengthening of Fe {\sc {ii}} and [Fe {\sc {ii}}]
lines compared with the August and December spectra; especially pronounced is
the Fe {\sc {ii}} $\lambda$4556 line, which increased its flux manifold.

\begin{table*}
\centering
\begin{minipage}{140mm}
\caption{Observed line properties in the spectrum of B416. \label{lines}}
\begin{tabular}{@{}llcc@{}}
Line & Source     & Flux (x$10^{-15}$ erg $s^{-1} cm^{-2}$) & EW (\AA) \\
H$\gamma$4340     & star   & 7.9$\pm$0.4   & -5.0$\pm$0.8 \\
                  & nebula & 0.9$\pm$0.3   & $---$ \\
He{\sc{i}}4471    & star   & 1.8$\pm$0.4   & -1.2$\pm$0.3 \\
He{\sc{i}}4713    & star   & 1.0$\pm$0.3   & -0.7$\pm$0.2 \\
H$\beta$4861      & star   & 25$\pm$2      & -17.5$\pm$0.4 \\
                  & nebula & 2.4$\pm$0.9   & $---$ \\
He{\sc{i}}4922    & star   & 1.4$\pm$0.2   & -1.0$\pm$0.1 \\
$[$O{\sc{iii}}$]$4959 & nebula & 0.22$\pm$0.02 & $---$ \\
$[$O{\sc{iii}}$]$5007 & nebula & 0.79$\pm$0.04 & $---$ \\
He{\sc{i}}5876    & star   & 4.6$\pm$1.5   & -6$\pm$2 \\
$[$N{\sc{ii}}$]$6548  & nebula & 0.20$\pm$0.04 & $---$ \\
H$\alpha$6563     & star   & 87$\pm$2      & -106$\pm$2 \\
                  & nebula & 3.3$\pm$0.4   & $---$ \\
$[$N{\sc{ii}}$]$6584  & nebula & 0.74$\pm$0.01 & $---$ \\
He{\sc{i}}6678    & star   & 1.70$\pm$0.03 & -3$\pm$1 \\
$[$S{\sc{ii}}$]$6717  & nebula & 0.5$\pm$0.2   & $---$ \\
$[$S{\sc{ii}}$]$6731  & nebula & 0.40$\pm$0.07 & $---$
\end{tabular}
\end{minipage}
\end{table*}

\subsubsection{Variations in the continuum \label{3.3.2}}

Our next goal was to see whether our spectral data portrays the
observed photometric variability described in \S \ref{3.1}. The red
continuum indeed shows no variation, which is in agreement with the
very low amplitude observed for the R and I LCs. As mentioned in \S
\ref{3.3.1}, the blue continuum tends to increase from August to
December. These continuum variations seem not to be related to the
8\fd26 periodicity, since according to the ephemeris (\S 3.2) the
calculated phases for the times of the APO spectra are: 0.58, 0.85 and
0.27 for the August, November and December images respectively. This
should imply a near-maximum light for August and a near-minimum light
for November, contrary to what is seen in Fig.~\ref{APOblue}.  Instead
we recall the fact that the LC of B416 displayed a gradual increase in
magnitude (\S \ref{3.1.2}) during the latest season. This light
increase was more prominent in B than in V and is also seen in the
spectra by the blue tail that grows gradually from August to December
1998. Therefore we suggest that our spectra display continuum
variations mainly due to the long-term variability of the star.

\subsubsection{Variations in the lines \label{3.3.3}}

Variations in emission lines are clearly seen for the He {\sc {i}}, Fe {\sc
{ii}} and [Fe {\sc {ii}}] lines (\S 3.3.1). Four of the He {\sc {i}}
lines (ruling out the He {\sc {i}}5876, recorded in the Red branch
of our spectrometer, which is noisier than
the Blue one) also show the development of P Cygni profiles in the November
spectrum (close to minimum phase) as described in \S 3.3.1. Due to the minor
contribution of the emission lines to the total flux, it is very unlikely that
the emission variation in the He and Fe lines alone is the source for the light
variation. Since we have only three spectra spread over some 100 days, it is
difficult to learn whether these variations are related to the short-
or long-term variations or any combination of the two.

\subsubsection{The nebula \label{3.3.4}}

B416 is surrounded by nebulosity as can be seen clearly in
Figure~\ref{nebula}. This nebulosity is No. 77 in the catalogue of
Court\`{e}s \& Cruvellier (1965) of H {\sc {ii}} regions in M33 and is
a relatively bright H {\sc {ii}} region in that galaxy. Boulesteix et
al. (1974) describe this H {\sc {ii}} region as having a physical size
of about 105 $\times$ 82 pc (equivalent to the angular size of 27 $\times$ 21
arcseconds on the sky) and a distance of about 1.2 kpc from the centre
of M33. The ring-shape structure of this H {\sc {ii}} region around
B416 is also seen in the spatial axis of the spectra and might hint
at a connection with the star. It therefore suggests that
B416 is a comparatively young system. The spectra of the line-of-sight
nebula (as described in \S {2.2}) show the characteristic nebular
lines, i.e.  Balmer, [O {\sc {iii}}], [S {\sc {ii}}] and [N {\sc {ii}}]
emission.  Measured fluxes of the nebular lines are listed in
Table~\ref{lines}.  As in the case of B416, the lines remain fixed at a
blueshifted system of $\sim$200 km/sec. The nebular lines are very
narrow and have practically no recognizable Doppler broadening in our
resolution. Due to the absence of [O {\sc {iii}}] $\lambda$4363 and [N
{\sc {ii}}] $\lambda$5755 lines from the spectrum, we are not able to make
a definite estimation of the temperature in the nebula. However, the
absence of the first line suggests low temperatures. On the other hand
the two [S {\sc {ii}}] $\lambda \lambda$6717,6731 lines enable us to
estimate the electron density in the nebula in the regular manner,
using the data presented in Table~\ref{lines}. The ratio [S {\sc {ii}}]
$\lambda$6717/$\lambda$6731 estimated as $\sim$1.3, implies an electron
density of N$_{e}$ $\sim$100 $cm^{-3}$. We also extracted the ratio
$N^{+} / S^{+}$ (the absence of [N {\sc {ii}}] $\lambda$5755 does not
change the ratio much due to its relatively low contribution) and
obtain a value of $\sim$1, which is characteristic of an H {\sc {ii}}
region rather than a circumstellar LBV shell \cite{pas98}.

The Balmer decrement of the nebular emission lines can be used, in
principle, for estimating the interstellar extinction in the direction
to B416. The H$\alpha / H\beta$ ratio has in our spectra a mean value
of $\sim$1.5. There is however a very large uncertainty in this value,
due to the fact that in our measurements of the two lines, two
different CCD chips are employed and the flux of both lines is very
low. The H$\gamma / H \beta$ ratio takes the value $\sim$0.4. Although
the uncertainty in this value is also large, it seems to indicate a
rather low extinction towards the H {\sc {ii}} region. This is
consistent with the general low value of the foreground extinction in
the direction of M33 \cite{mas95}.

\section{Discussion \label{4}}

In this section we discuss the implications of the observational results
presented in this work on our understanding of the nature of B416, on its
proper evolutionary status and on some of its observational properties.

\subsection{The nature of the periodicity \label{4.1}}

One of the most common astronomical clocks that produce observed coherent
periodic variations is the orbital revolution of a stellar binary system.
In fact, the structure of the periodic component in the LC of B416, presented
in Fig.~\ref{harmony}  is reminiscent of an eclipsing binary light curve. This
possibility is however severely constrained by the photometric properties of
B416 that we found, namely, the colour-dependence of the periodic variation and
the 7\% change in total flux between minimum and maximum phase. From the values
given in \S 1.2 it is apparent that the system's luminosity is at least
3x10$^5$L$_{\sun}$, hence the dip in the LC corresponds to about
2x10$^4$L$_{\sun}$. Two leading eclipsing models are considered: (1) a
small blue star with luminosity of 2x10$^4$L$_{\sun}$ being occulted by
the brighter primary; (2) a rather faint star (less than
2x10$^4$L$_{\sun}$) that can cover in its inferior conjunction an area
of roughly 7\% of the surface of the primary.  Both models fail
primarily on the basis of not being able to explain the colour-dependent
variation. Spectroscopically, there are no detectable Doppler shifts
larger than $\sim$20 km/sec in the lines and no absorption lines or
bands are observed, which might account for a companion.  These
spectral properties also set serious limitations on the mass and luminosity
of the companion. For example, if we assume that the mass of the
primary is of the order of 20M$_{\sun}$ and that the inclination angle
of the system is 90\degr \ then a velocity of less than 20 km/sec
corresponds to a mass of less than 1.5M$_{\sun}$ for the companion,
which in turn cannot account for the observed dip in brightness if it
is a main sequence star. On the other hand, we can assume that the
companion may be a compact object, surrounded by an accretion disc
which contributes the $\sim$10$^4$L$_{\sun}$ energy and therefore its
occultation can be the source of variation. Such an energy output by an
accretion disc is not observed even for high mass X-ray binaries. A
white dwarf or a neutron star companion hypothesis can be rejected also on the
grounds of evolution, since we hardly expect to find an evolved star inside a
young H {\sc{ii}} region orbiting a very young and massive system like
B416. Even if the progenitor of the evolved star lost most of its mass
to the primary, it is unlikely that several 10M$_{\sun}$ were
transfered within the lifetime of the surrounding nebula. The likelihood of a
neutron star or a black hole, two possible relics of a supernova, to be a
companion of B416 is also small since there is no observational indication,
photometrically or spectroscopically of the presence of a SNR in the area.
We therefore regard binarity as a highly unlikely cause of the observed
periodic light variation.

\subsection{Single star interpretation (LBV?)\label{4.2}}

Rejecting binarity as the source for the periodic variation does not mean
however, that B416 is not a binary or even a multiple system, since it may be a
non-eclipsing, low inclination binary system. We proceed, however, with our
classification of B416, considering it as a single star. This is justified
because it appears that the observed high luminosity and most of the stellar
spectral features are indeed contributed by one star alone. In order to do so
we list here the observed major spectroscopic and photometric properties of
B416:

\begin{enumerate}
\renewcommand{\theenumi}{(\arabic{enumi})}
\item M$_{V}=-8.8$
\item B$-$V $\sim$0
\item U$-$B $\sim -$0.8
\item L$_{{\rm H}\alpha} \sim 1700 L_{\sun}$
\item Large amplitude light variations on long time scales
\item Light variations ($\leq$0.1 mag) on a short time scale
\item Amplitude of light variations increases bluewards
\item Spectrum shows H {\sc{i}}, He {\sc{i}}, Fe {\sc{ii}} and [Fe {\sc{ii}}]
      emission
\item Spectral variations in both continuum and lines
\item P Cygni profiles exhibiting velocities of $\sim$300 km/sec
\end{enumerate}

The first property listed above is taken from Massey et al. (1996), who
also give the apparent V magnitude of B416 as 16.29 (consistent with
the value given by Humphreys \& Sandage (1980), see \S 1.2). Taking
into account the estimated extinction due to the foreground reddening
of M33 A$_V$=0.22 and a distance modulus of $\mu$=24.5 \cite{vdb91}, we
obtain an estimated absolute magnitude of M$_V=-$8.43. This value is
different by some 0.4 magnitudes from the value given by Massey et al.
(1996). The source of the difference is probably in a different distance
modulus and/or a different extinction value that they used. If we assume
even smaller or no extinction at all towards B416, as we found for the
surrounding nebula in \S 3.3.4, the value of M$_V$ may increase up to
M$_V=-$8.21. But even this number indicates a very luminous star. While the
photometric characteristics refer to an A-type supergiant, the spectrum does
not resemble that of an A-type star, nor that of any O- or B-type stars,
particularly since it does not show any absorption features.  The spectrum of
B416 also does not resemble that of a WR star nor of an Ofpe/WN9 star
(reclassified as WN9-11 stars by Smith, Crowther and Prinja (1994)), mainly due
to the absence of He {\sc{ii}} and N {\sc{iii}} lines. The spectrum of B416
rather indicates relatively low temperatures, consistent with the measured
intrinsic zero B$-$V colour. Given its comparatively high luminosity, its
variational properties and the peculiar spectral features as listed above, we
are bound to classify B416 as an LBV.

It was already mentioned in \S 1.2 that on the basis of total
luminosity, colour and spectrum alone, B416 can in fact be classified only as
an LBV candidate, in quiescence. According to our study, B416 indeed shows
spectral resemblance with the spectrum given by Massey et al. (1996),
who also emphasizes the fact that spectroscopically B416 and the other four
newly found LBV candidates in M33, all resemble each other and also the star
Var C, the well known LBV in M33. The spectrum of B416 taken by Massey et
al. (1996), which shows the characteristic LBV quiescence spectrum, was
taken in 1995 (just three years before ours) and the relatively high UV
flux they measure for the star further confirms the warm phase scenario \cite{hum94}. The appearance of prominent He {\sc{i}} emission lines in our
spectra is consistent with this warmer phase of B416, i.e.  visual minimum
state. Our classification receives yet another support from the latest work on
the discovery of LBV candidates in M31 \cite{kin98}. All five candidates
discovered in that galaxy show remarkable spectral similarity to B416. In
particular we mention the observed properties of B416 that can be
deduced from Table~\ref{lines} and \S 3.3 above, such as the [S
{\sc{ii}}]/H$\alpha$ ratio, the H$\alpha$ luminosity (which is believed
to be contributed mainly by the stellar wind), EW(H$\alpha$) and the
terminal velocity. All these are quite in agreement with the values given in
that work for the LBV candidates in M31.

As an LBV, B416 is certainly among the faintest members of this class of stars.
We may sub-classify it as an R 71 type LBV \cite{boh97} and
locate it at the bottom of the S-Dor instability strip in the HR
diagram, among R 71 of the LMC, R 40 of the SMC and HR Car of our own
galaxy \cite{hum94}. This suggested location on the HR diagram implies
that at its current minimum state, the effective temperature of B416 is
of the order of T$\sim$12,000 K. As mentioned above, however, we are aware of
the fact that classification cannot be made on spectroscopic grounds alone,
even when it comes to such rare objects as LBVs.

In order to be fully entitled as an LBV, a star has to change its
magnitude considerably, by some 0.5 to 2 magnitudes, on a time scale of
years or decades \cite{boh97}. B416 was not observed or reported to
have large light variations spanning years or decades, and in this
sense it could undermine our classification. If, however, an
amplitude-luminosity relation exists for LBVs, as suggested by
Humphreys and Davidson (1994), then B416 has potentially quite a small
amplitude for such long-term variations. These variations, commonly
referred to as S-Dor (SD) phases \cite{ste97}, might be very small for
B416.  They might well have been overlooked in the past, since little
or any photometric data, beside those in this work, exist for the
star.  It may well also be that the 0.17 magnitude difference in Clear
that we measured between 1987 and 1997 is indicative of a small scale
SD phase of B416, as expected in low luminosity LBVs. On the other
hand, this difference in magnitude can be just a fraction of a larger
amplitude SD phase, that might have occured between those two years
or even before 1987. We therefore believe that it is suitable to
classify B416 as a true LBV.

\subsection{Periodic microvariations in B416\label{4.3}}

In the previous section we argued that according to our observations
and the available archived data, it is very likely that B416 is a
relatively faint LBV with a mass of the order of 20M$_{\sun}$
\cite{hum94}. From here it is a straightforward step to describe the
small amplitude light variations that we observe as the microvariations
of this LBV star. The period $\sim$8 day is well within the range of
quasi-periods measured in known LBVs (see Breysacher 1997 and Lamers et
al. 1998). These microvariations, also observed in $\alpha$ Cygni type
supergiants, are mainly irregular or quasi-periodic \cite{lam98} and
are attributed to pulsational instabilities of the LBV (or blue
supergiant) due to the increase of luminosity up to and near the
Eddington luminosity. The observed microvariations in B416 increase in
amplitude with decreasing wavelength.  This behaviour is observed in
LBVs in quiescence, since during that stage of the SD phase the star is
more compact and hot, giving it a blue colour \cite{lam98}. Periodic
microvariations were recently reported in the LBV HD 5980
\cite{bre97}.  This detection was based on observations that were
carried out during one observational season (between 1995 and 1996).
Our data indicate periodic microvariations of B416, which are coherent
and stable over a whole decade.






\section{Summary}

We conclude that B416 is an LBV of the rather fainter sub-class R 71.
It should therefore be added to the list of 32 known LBVs \cite{boh97}
as the fifth known LBV in M33 and could be placed near R 71 on the HR
diagram.  The detected periodic light variation is evidence for
periodic and stable microvariations found for the first time in an
LBV.  A continued photometric followup of this interesting star is
required in order to look for the anticipated SD phase and to keep an
eye on the clock of the microvariations. A dense spectroscopic
monitoring is also required in order to correlate the spectra with the
microvariations, while high resolution spectroscopy might detect the
LBV circumstellar nebula, which is expected around LBVs due to a large
amount of mass loss \cite{hum94}.

\section*{Acknowledgments}

We would like to thank Noah Brosch who helped in the coordination of
the initial 1986/87 Wise Observatory photometric survey of luminous
stars in M33, and Anna Heller who participated in it and performed most
of its data reduction. We acknowledge the assistance of Gaghik
Tovmassian and Camron Hastings in obtaining some of the APO spectra. We
also thank Shai Kaspi and Liliana Formiggini for their assistance in
the photometric and spectroscopic reduction procedures. We are grateful
to WO staff members Ezra Mashal, Friedel Loinger, Sami Ben-Gigi and
John Dan for their crucial contribution to this project. This research
has made use of the SIMBAD database, operated at CDS, France. Astronomy
at the WO is supported by grants from the Israel Science Foundation.
We would like to thank the referee for his useful comments.

\label{lastpage}

\end{document}